\documentclass[aps,prd,preprint,groupedaddress]{revtex4}

\usepackage{graphicx}
\usepackage{amsfonts}
\usepackage{epsfig}
\usepackage{amssymb}
\usepackage{amsmath}
\usepackage{epstopdf}
\newcommand{\be}{\begin{eqnarray}}
\newcommand{\ee}{\end{eqnarray}}
\newcommand{\bi}{\begin{itemize}}
\newcommand{\ei}{\end{itemize}}
\newcommand{\Tr}{{\rm Tr}}
\newcommand{\Rea}{{\rm Re}}
\newcommand{\Ima}{{\rm Im}}

\newcommand{\Log}{{\rm Log}}

\begin{document}

\title{Standard Model CP-violation and \\Cold Electroweak Baryogenesis}
\author{Anders Tranberg}
\email[]{anders.tranberg@helsinki.fi}
\affiliation{Helsinki Institute of Physics, P.O.Box 41, FIN-00014 Helsinki, Finland and \\
  Department of Physical Sciences, University of Oulu, FIN-90014 Oulu, Finland}
\date{\today}

\begin{abstract}
Using large-scale real-time lattice simulations, we calculate the baryon asymmetry generated at a fast, cold electroweak symmetry breaking transition. CP-violation is provided by the leading effective bosonic term resulting from integrating out the fermions in the Minimal Standard Model at zero temperature, and performing a covariant gradient expansion \cite{Hernandez:2008db}. This is an extension of the work presented in \cite{Tranberg:2009de}. The numerical implementation is described in detail, and we address issues specifically related to using this CP-violating term in the context of Cold Electroweak Baryogenesis. The results support the conclusion of \cite{Tranberg:2009de}, that Standard Model CP-violation may be able to reproduce the observed baryon asymmetry in the Universe in the context of Cold Electroweak Baryogenesis.
\end{abstract}


\maketitle


\section{Introduction\label{sec:introduction}}

At its inception, Electroweak Baryogenesis was an attempt to explain the observed baryon asymmetry of the Universe by processes originating from Standard Model physics only \cite{Kuzmin:1985mm,Rubakov:1996vz}. Although essential ingredients, C-, P-, CP- and baryon number violation are present in the Standard Model, the detailed quantitative implementation of the scenario has encountered two stumbling blocks: At experimentally allowed Higgs masses, the electroweak transition is an equilibrium cross-over \cite{Kajantie:1996mn}, rather than the required out-of-equilibrium phase transition; and at electroweak-scale temperatures, the effective CP-violation is much too small to produce the observed asymmetry \cite{Shaposhnikov:1987pf,Gavela:1994ds,Gavela:1994dt}.

The way to alleviate these problems has traditionally been to embed the Standard Model in a larger theory, either by extending the Higgs sector (see, for instance \cite{Fromme:2006cm}), or by imposing full-fledged supersymmetry (see \cite{Carena:2002ss,Prokopec:2003is,Carena:2004ha,Konstandin:2005cd,Huber:2006wf,Chung:2008aya} for recent developments). This allows for a first order electroweak phase transition, while introducing additional coupling constants in the Higgs sector, which if taken complex may give rise to in principle an arbitrary amount of CP-violation.

Standard Model CP-violation originates from the CKM-mixing in the fermion mass matrix, but is often phrased in terms of effective bosonic terms in the action, appearing as a result of integrating out the fermions in the path integral. The strength of CP-violation is encoded in the coefficient of the CP-violating terms, and is proportional to 
\be
 J\left(m^2_t-m^2_c\right)\left(m_t^2-m_u^2\right)\left(m^2_c-m_u^2\right)\left(m_b^2-m_s^2\right)\left(m_b^2-m_d^2\right)\left(m_s^2-m_d^2\right)
\ee
in terms of the quark masses $m_i$ and the Jarlskog invariant \cite{Jarlskog:1985ht},
\be
J\simeq 3.1\times 10^{-5}.
\ee
At high temperatures $T>m_i$, the coefficient is known to be very small $\propto J(m_i/T)^{12}$ \cite{Shaposhnikov:1987pf}, and electroweak baryogenesis at these temperatures require additional sources of CP-violation. 

At zero temperature, the situation changes radically. In \cite{Salcedo:2000hp,Salcedo:2000hx,Konstandin:2003dx,Hernandez:2007ng,Smit:2004kh,GarciaRecio:2009zp} two separate methods have been developed to calculate the CP-violating part of the effective bosonic action at zero temperature. Integrating out the fermions, we have
\be
Z[\Psi,\bar \Psi,\phi,A]=\int \mathcal{D}\phi\mathcal{D}A\mathcal{D}\Psi\mathcal{D}\bar\Psi e^{iS[\Psi,\bar \Psi,\phi,A]}
=\int \mathcal{D}\phi\mathcal{D}A e^{iS[\phi,A]+i\Tr\Log D[\phi,A]}
\ee
with $D$ the fermion propagator and $\phi$ and $A$ the Higgs and gauge fields, respectively. In both approaches, the fermion contribution is then expanded in covariant gradients\footnote{Powers of $D_\mu\phi$ and $A_\mu$.}. At leading order (order 4 in derivatives), CP-violation is absent \cite{Smit:2004kh}, but at next-to-leading order (order 6), CP-violating terms have been reported in the P-odd \cite{Hernandez:2008db} and P-even \cite{GarciaRecio:2009zp} sectors. The authors of \cite{GarciaRecio:2009zp} also finds vanishing CP-violation at order 6 in the P-odd sector in conflict with \cite{Hernandez:2008db}.

The crucial point is that at zero temperature the coefficients of these terms are only suppressed by $J$, but not by powers of $m_i/v$, $v$ the Higgs vev, as one may have expected. This opens up the possibility that Standard Model CP-violation may be responsible for the baryon asymmetry, if baryogenesis took place at low enough temperature. 

Cold Electroweak Baryogenesis is realized if the temperature is well below the electroweak scale when the electroweak transition happens \cite{GarciaBellido:1999sv,Krauss:1999ng,Copeland:2001qw,Turok:1990in} (further developed in \cite{Tranberg:2003gi,vanTent:2004rc,vanderMeulen:2005sp,Tranberg:2006ip,Tranberg:2006dg}). In the Standard Big Bang scenario, inflation and reheating occur at very high temperature, and in the Standard Model the electroweak cross-over takes place around $T=100\,$GeV. However, if Higgs symmetry breaking is triggered not by the change in the finite-temperature potential but by the coupling to another field, the transition can be delayed until cosmological expansion has cooled the Universe far below the electroweak scale. Different mechanisms to produce this transition have been proposed including low-scale hybrid inflation \cite{Copeland:2001qw,vanTent:2004rc}, two-step symmetry breaking \cite{Enqvist:2010fd} and a specific variant of a first order phase transition \cite{Konstandin:2011ds}.
 
In terms of the Higgs potential, we write
\be
V(\phi)=\mu_{\rm eff}^2(t)\phi^\dagger\phi+\lambda(\phi^\dagger\phi)^2,
\ee
with the electroweak transition happening when $\mu^2_{\rm eff}(t)=0$. For the purpose of the present work, all we need is that the Higgs mass parameter flips sign quickly, so that to a good approximation
\be
\mu_{\rm eff}^2(t<0)=\mu^2, \qquad \mu_{\rm eff}^2(t>0)=-\mu^2,
\ee
i.e. an instantaneous quench. The system goes through a spinodal (or tachyonic) instability, where field modes with $k<\mu$ grow exponentially until Higgs self-interactions become significant \cite{Felder:2000hj}. This stage of (p)reheating leads to large field occupation numbers in the IR \cite{Rajantie:2000nj,GarciaBellido:2002aj,Smit:2002yg,GarciaBellido:2003wd,Skullerud:2003ki,DiazGil:2007dy}, making the dynamics effectively classical \cite{GarciaBellido:2002aj,Smit:2002yg,GarciaBellido:2003wd}.

In the following, we will assume that Standard Model CP-violation manifests itself at leading order as the CP-odd term of \cite{Hernandez:2008db}, with some coefficient $\kappa^{\rm CP}$, and that the asymmetry scales linearly in $\kappa^{\rm CP}$ (this is certainly true for small values of $\kappa^{\rm CP}$). In numerical simulations of a cold electroweak transition, we will calculate the baryon asymmetry generated for a given $\kappa^{\rm CP}$. This will allow us to estimate a value of $\kappa^{\rm CP}$, above which Standard Model CP-violation can accommodate the observed asymmetry. Awaiting the resolution of the discrepancy in \cite{Hernandez:2008db,GarciaRecio:2009zp}, we will consider the specific CP-violating term used here as representative of all P-even and P-odd terms at NLO in the covariant gradient expansion. The important and common feature to both results is that the coefficients of these terms are not suppressed by powers of the Yukawa couplings ($m_i/v$). 

We will follow the dynamics of the system starting from a vacuum ($T=0$) initial state, and calculate the baryon number asymmetry produced in the transition under the influence of the CP-violating term. The implementation closely mirrors the work in \cite{Ambjorn:1990pu,Tranberg:2006ip,Tranberg:2006dg}, except that the CP-violating term is different (and much more complicated).

Preliminary results were presented in \cite{Tranberg:2009de}, where we found that Standard Model CP-violation as represented by the leading bosonic term can indeed account for the observed baryon asymmetry. The present paper is an update confirming this conclusion with siginficantly more statistics and providing the numerical details of the simulations.


\section{SU(2)-Higgs model with CP-violation\label{sec:su2higgs}}

Serving as an approximation to the full Standard Model, we consider the SU(2)-Higgs model with CP-violation, with continuum action
\be
-\int\,d^4x\left[\frac{1}{2g^2}\Tr F_{\mu\nu}F^{\mu\nu}+\left(D_\mu\phi\right)^{\dagger}D^\mu\phi+\mu_{\rm eff}^2(t)\phi^{\dagger}\phi+\lambda\left(\phi^{\dagger}\phi\right)^2+V_0-S_{\rm CP, 6}\right],
\ee
in terms of a complex scalar doublet and an SU(2) gauge field
\be
\phi(x)=\left( \begin{array}{c}\phi_1(x)\\\phi_2(x)\end{array} \right),\qquad
A_\mu(x)=A_\mu^a(x)\frac{\sigma^a}{2},
\ee
where $\sigma^a$ are the Pauli matrices. We have $F_{\mu\nu}=\partial_\mu A_\nu-\partial_\nu A_\mu-i\left[A_\mu,A_\nu\right]$, $D_\mu\phi=\left(\partial_\mu-iA_\mu\right)\phi$. The vacuum particle masses are $m_W=gv/2$, $m_H=\sqrt{2}\mu=\sqrt{2\lambda}v$ in terms of the Higgs vacuum expectation value $v=\mu/\sqrt{\lambda}$, with $V_0=\mu^4/(4\lambda)$ normalising the energy to zero in the vacuum.

We also define the useful
\be
e^+=\frac{\sigma^1+i\sigma^2}{2}
,\quad e^-=\frac{\sigma^1-i\sigma^2}{2}
,\quad e^0=-\frac{\sigma^3}{2},
\ee
and note that
\be
\left(e^\pm\right)^\dagger=\left(e^\pm\right)^T=e^\mp,\quad  \left(e^0\right)^\dagger=\left(e^0\right)^T=e^0.
\ee
We want to write everything in terms of the Higgs matrices, with\footnote{Note that we use the $\phi$ upside down compared to the notation in \cite{Hernandez:2008db}}
\be
\Phi_x=\left( \begin{array}{cc}
\tilde{\phi}_1(x) & \phi_1(x)\\
\tilde{\phi}_2(x) & \phi_2(x)   
\end{array} \right)
=
\left( \begin{array}{cc}
\phi_2^*(x) & \phi_1(x)\\
-\phi_1^*(x) & \phi_2(x)   
\end{array} \right),
\ee
or
\be
\tilde{\phi}(x)=i\sigma^2\phi^*(x).
\ee
We have
\be
|\phi^2|=\frac{1}{2}\Tr\, \Phi^\dagger\Phi,
\ee
and for some matrix B
\be
\Phi^\dagger(x) B(x,y)\Phi(y)=\left( \begin{array}{cc}
\tilde{\phi}^\dagger(x)B(x,y)\tilde{\phi}(y)& \tilde{\phi}^\dagger(x)B(x,y)\phi(y)\\
\phi^\dagger(x)B(x,y)\tilde{\phi}(y)& \phi^\dagger(x)B(x,y)\phi(y)  
\end{array} \right).
\ee
We then define
\be
B^0&=&\Tr e^0 \Phi^\dagger(x) B(x,y)\Phi(y)=\phi^\dagger(x)B(x,y)\phi(y)-\tilde{\phi}^\dagger(x)B(x,y)\tilde{\phi}(y),\nonumber\\
&&\\
B^+&=&\Tr e^+ \Phi^\dagger(x) B(x,y)\Phi(y)=\phi^\dagger(x)B(x,y)\tilde{\phi}(y),\\
B^-&=&\Tr e^- \Phi^\dagger(x) B(x,y)\Phi(y)=\tilde{\phi}^\dagger(x)B(x,y)\phi(y).
\ee
We are particularly interested in the case when $B$ is composed of covariant derivatives $D$, and we define the following:
\be
B&\rightarrow& D_\mu(x,y):\quad C^{\pm,0}_\mu=\Tr\, e^{\pm,0} \Phi^\dagger(y) D_\mu(x,y)\Phi(x),\\
B&\rightarrow& D_\nu(x,y)D_\lambda(x,y):\quad D^{\pm}_{\nu\lambda}=\Tr\, e^{\pm} \Phi^\dagger(y) D_\nu(x,y)D_\lambda(x,y)\Phi(x).
\ee
We note the important relations under complex conjugation
\be
\left(C^\pm_\mu\right)^*=-C^\mp_\mu,\qquad \left(C^0_\mu\right)^*=-C^0_\mu,\qquad \left(\epsilon^{\mu\nu\lambda\sigma}D^\pm_{\nu\lambda}\right)^*=-\epsilon^{\mu\nu\lambda\sigma}D^\mp_{\nu\lambda},
\ee
which follow from the definitions of the covariant derivative,
\be
\epsilon^{\mu\nu\lambda\sigma}F^{\nu\lambda}=\epsilon^{\mu\nu\lambda\sigma}i[D_\nu,D_\lambda]=2i\epsilon^{\mu\nu\lambda\sigma}D_\nu D_\lambda,
\ee
The CP-violating term can be written \cite{Hernandez:2008db} as
\be
S_{\rm CP, 6}=\int d^4x\,\frac{i\kappa}{\left(\phi^2_\Lambda\right)^6}\epsilon^{\mu\nu\lambda\sigma}\bigg(C_\mu^0\bigg(C^+_\sigma C^-_\alpha C^-_\alpha+C^-_\sigma C^+_\alpha C^-_\alpha\bigg)D_{\nu\lambda}^+\nonumber\\
+C_\mu^0\bigg(C^-_\sigma C^+_\alpha C^+_\alpha+C^+_\sigma C^-_\alpha C^+_\alpha\bigg)D_{\nu\lambda}^-\bigg).\nonumber\\
\ee
where $\phi^2_\Lambda$ will be defined below. Note that the two terms are each other's complex conjugates, but with a minus sign\footnote{Actually one for each factor, so $(-1)^5$.}.
Using $v=246\,$GeV, $m_W=80.4\,$GeV and $\tilde{m}_c=1.3\,$GeV, $J=3.1\times 10^{-5}$, we have
\be
\kappa \simeq 1.3\times 10^{-4}\kappa^{\rm CP}.
\ee
The Standard Model, zero temperature, value for $\kappa^{\rm CP}$ was found in \cite{Hernandez:2008db} to be 9.87.

\subsection{The cut-off  $\Lambda$\label{sec:cut-off}}

The CP-violating term is divergent at points in space where $\phi^\dagger\phi=0$, where the gradient expansion breaks down. Although rare once the electroweak transition has begun, at any given time during the simulation, there will exist small regions of space where the CP-violating force is ``unphysically'' large. Apart from discarding the approach altogether, the solution is to identify these regions and cut them out of the CP-violating dynamics. This can for instance be done by imposing a threshold $\Lambda$, and set the CP-violating force to zero whenever $\sqrt{\phi^\dagger\phi(x)}<\Lambda$. Rather than a step function, in practice we will we introduce the cut-off Higgs field $\phi^2_\Lambda$
\be
\label{eq:phi_lambda}
\frac{1}{\phi^\dagger\phi}\quad \rightarrow \quad \frac{1}{\phi^2_\Lambda}=\frac{1}{c\left(\phi^\dagger\phi+\Lambda^2\right)},
\ee
fixing the number $c$ by
\be
c=\frac{1}{1+\frac{2\Lambda^2}{v^2}},
\ee
so that 
\be
\phi^\dagger\phi=\frac{v^2}{2} \rightarrow \frac{1}{c\left(\phi^\dagger\phi+\Lambda^2\right)}=\frac{1}{v^2/2}.
\ee
We emphasize that the cut-off $\Lambda$ is {\it in no way} related to a momentum cut-off. It is simply a threshold which moderates the divergence in the CP-violating force near small $|\phi(x)|$. We also note that the introduction of such a threshold means that the result of the simulations will be a {\it lower bound} on the generated asymmetry from the Standard Model, and there will be a dependence of the result on $\Lambda$. In \cite{Tranberg:2009de} we used $\Lambda=50\,$GeV. Below, we will study in detail the dependence on $\Lambda$. We note that $\Lambda\in [0,v/\sqrt{2}]=[0,174]\,$GeV.

\section{Lattice implementation\label{sec:lattice}}

We here explicitly write down the lattice action and equations of motion for including CP-violation. This runs along the lines of \cite{Ambjorn:1990pu,Tranberg:2006ip,Tranberg:2006dg}, and may be skipped by readers who are not interested in the numerical implementation.

On the lattice, we perform a rescaling of the Higgs field
\be
\Phi\rightarrow \sqrt{\lambda} a_x\Phi, 
\ee
We use the usual lattice derivatives
\be
a_\mu D_{\mu}\Phi_x=U_{x,\mu}\Phi_{x+\mu}-\Phi_x,\quad
a_\mu D'_{\mu}\Phi_x=\Phi_x-U_{\mu,x-\mu}^\dagger\Phi_{x-\mu}.
\ee
The symmetrized derivatives are
\be
D^s_\mu=\frac{1}{2}\left(D_\mu+D_\mu'\right),\qquad D^2_{\mu\nu}=D^s_\mu D^s_\nu.
\ee
Then let us define a slightly different lattice version of our factors for $\mu\ne 0$
\be
\phi^2_\Lambda(x)C_\mu^i(x)=2a_\mu\Tr[e^i\Phi^\dagger_x D^s_\mu \Phi_{x}]= \Tr [e^i\Phi^\dagger_x \left(U_{\mu,x}\Phi_{x+\mu}-U_{\mu,x-\mu}^\dagger\Phi_{x-\mu}\right)],\quad i=+,-,0,\nonumber\\
\ee
and a special version for the timelike factor $\mu=0$,
\be
\phi^2_\Lambda(x)C_0^i(x)&=&\Tr[e^i\left(\Phi^\dagger_x D_0 \Phi_{x}+\Phi^\dagger_{x+0} D_0' \Phi_{x+0}\right)]\nonumber\\&= &\Tr [e^i\left(\Phi^\dagger_x (U_{0,x}\Phi_{x+0}-\Phi_{x})+\Phi^\dagger_{x+0}(\Phi_{x+0}-U_{0,x}^\dagger\Phi_{x})\right)],\quad i=+,-,0.\nonumber\\
\ee
This is necessary to avoid the equations of motion from becoming implicit {\it two} steps ahead in time. In this way they are implicit {\it one} step only.
We note, that
\be
\Tr [e^i \Phi^\dagger_x\Phi_x]=0,\quad i=+,-,0.
\ee
We also define, for $\mu,\nu \neq 0$
\be
&&\phi^2_\Lambda(x)D_{\mu\nu}^\pm(x)=\Tr \Big[e^\pm\Phi^\dagger_x \bigg(U_{\mu,x}U_{\nu,x+\mu}\Phi_{x+\mu+\nu}+U^\dagger_{\mu,x-\mu}U^\dagger_{\nu,x-\mu-\nu}\Phi_{x-\mu-\nu}\nonumber\\&&-U^\dagger_{\mu,x-\mu}U_{\nu,x-\mu}\Phi_{x-\mu+\nu}-U_{\mu,x}U^\dagger_{\nu,x+\mu-\nu}\Phi_{x+\mu-\nu}\bigg)\Big].
\ee
Again, special rules apply for the timelike derivatives
\be
&&\phi^2_\Lambda(x)D_{0\nu}^\pm=\Tr \Big[e^\pm\left(\Phi^\dagger_x (U_{0,x}U_{\nu,x+0}\Phi_{x+0+\nu}-U_{\nu,x}\Phi_{x+\nu}-U_{0,x}U^\dagger_{\nu,x-\nu+0}\Phi_{x-\nu+0}+U^\dagger_{\nu,x-\nu}\Phi_{x-\nu})\right)\nonumber\\
&&+\left(\Phi^\dagger_{x+0}(U_{\nu,x+0}\Phi_{x+\nu+0}-U^\dagger_{0,x}U_{\nu,x}\Phi_{x+\nu}-U^\dagger_{\nu,x+0-\nu}\Phi_{x+0-\nu}+U^\dagger_{0,x}U^\dagger_{\nu,x-\nu}\Phi_{x-\nu})\right)\Big],\nonumber\\
\ee
and
\be
&&\phi^2_\Lambda(x)D_{\nu 0}^\pm=\Tr \Big[e^\pm\left(\Phi^\dagger_x (U_{\nu,x}U_{0,x+\nu}\Phi_{x+0+\nu}-U_{n,x}\Phi_{x+\nu}-U_{\nu,x-\nu}^\dagger U_{0,x-\nu}\Phi_{x-\nu+0}+U_{\nu,x-\nu}^\dagger\Phi_{x-\nu})\right)\nonumber\\
&&+\left(\Phi^\dagger_{x+0}(U_{\nu, x+0}\Phi_{x+\nu+0}-U_{\nu, x+0}U^\dagger_{0,x+\nu}\Phi_{x+\nu}-U_{\nu,x-\nu+0}^\dagger\Phi_{x-\nu+0}+U^\dagger_{\nu,x-\nu+0}U^\dagger_{0,x-\nu}\Phi_{x-\nu}\right)\Big].\nonumber\\
\ee
We note that compared to our continuum notation we have the substitution rules
\be
\phi^2_\Lambda(x)(C^i_\mu)^{lattice}&\quad& \leftrightarrow \quad 2a_\mu(C^i_\mu)^{continuum},\\
\phi^2_\Lambda(x)(D^i_{\nu\lambda})^{lattice}&\quad& \leftrightarrow \quad 4a_\nu a_\lambda(D^i_{\nu\lambda})^{continuum}.
\ee
Then the lattice contribution to the action is
\be
\label{eq:CP_8}
S_{\rm CP, 6}&=&\sum_{x,t}\epsilon^{\mu\nu\lambda\sigma}\frac{i\beta_k}{\phi^{2}_{\Lambda}} \Bigg(C_{\mu,x}^0 D_{\nu\lambda,x}^+\left(C_{\sigma,x}^+C_{\alpha,x}^-C_{\alpha,x}^-+C_{\sigma,x}^-C_{\alpha,x}^+C_{\alpha,x}^-\right)\frac{a_x^2}{a_{\alpha}^2}\nonumber\\&&+C_{\mu,x}^0 D_{\nu\lambda,x}^-\left(C_{\sigma,x}^-C_{\alpha,x}^+C_{\alpha,x}^++C_{\sigma,x}^+C_{\alpha,x}^-C_{\alpha,x}^+\right)\frac{a_x^2}{a_{\alpha}^2}
\Bigg)
\ee
with the dimensionless
\be
\beta_\kappa=\frac{3 J\kappa^{CP}(a_x^2m_H^2/4)}{2^{16}\pi^2a_x^2m_c^2}.
\ee
In the following we will use the shorthand
\be
\label{eq:betakappa}
\bar{\epsilon^{\mu\nu\lambda\sigma}}=\epsilon^{\mu\nu\lambda\sigma}\frac{a_x^2}{a_{\alpha}^2}.
\ee
The complete lattice action of the SU(2)-Higgs model with CP-violation then reads,
\be
S=\sum_{x,t}\Big[&&+\beta_G^t\sum_n\left(1-\frac{1}{2}\Tr[U_{x,0}U_{x+0,n}U^\dagger_{x+n,0}U^\dagger_{x,n}]\right)\nonumber\\
&&-\beta_G^s\sum_{m<n}\left(1-\frac{1}{2}\Tr[U_{x,m}U_{x+m,n}U^\dagger_{x+n,m}U^\dagger_{x,n}]\right)\nonumber\\
&&+\beta_H^t\frac{1}{2}\Tr[\left(U_{0,x}\Phi_{x+0}-\Phi(x)\right)^\dagger\left(U_{0,x}\Phi_{x+0}-\Phi(x)\right)]\nonumber\\
&&-\beta_H^s\sum_n\frac{1}{2}\Tr[\left(U_{n,x}\Phi_{x+n}-\Phi(x)\right)^\dagger\left(U_{n,x}\Phi_{x+n}-\Phi(x)\right)]\nonumber\\
&&-\beta_R\left(\frac{1}{2}\Tr\Phi^\dagger_{x}\Phi_x-v_{\rm lat}^2\right)^2\Big]-S_{\rm CP, 6},
\ee 
where by matching to the continuum theory, we have in addition to (\ref{eq:betakappa})
\be
\beta_G^t=\frac{4}{g^2}\frac{a_x}{a_t},\quad \beta_G^s=\frac{4}{g^2}\frac{a_t}{a_x},\quad  \beta_H^t=\frac{1}{\lambda}\frac{a_x}{a_t},\quad \beta_H^s=\frac{1}{\lambda}\frac{a_t}{a_x},\quad \beta_R=\frac{1}{\lambda}\frac{a_t}{a_x},\quad v^2_{\rm lat}=\frac{(a_xm_H)^2}{4}.\nonumber\\
\ee


\subsection{Higgs equation of motion\label{sec:higgseom}}

The Higgs equation of motion, in the $A_0=0$ gauge, reads
\be
\partial_0'\partial_0\Phi_y=\frac{\beta_H^s}{\beta_H^t}\sum_n \left(U_{n,y}\Phi_{y+n}+U^\dagger_{n,y-n}\Phi_{y-n}-2\Phi_{y}\right)-\frac{2\beta_R}{\beta_H^t}\left(\frac{1}{2}\Tr[\Phi^\dagger_y\Phi_y]-v^2_{\rm lat}\right)\Phi_y-\frac{1}{\beta_H^t}\frac{\delta S_{\rm CP, 6}}{\delta\Phi_y^\dagger},\nonumber\\
\ee
with
\be
\label{eq:Heom}
&&\frac{\delta S_{\rm CP, 6}}{\delta\Phi^\dagger_y}=\bar{\epsilon^{\mu\nu\lambda\sigma}}\frac{\beta_\kappa c}{\phi^2_\Lambda(y)}\Big(\frac{6}{\phi_c^2(y)}\times 2\Ima(c_\phi(y))\Phi_y
\nonumber\\
&&-2\delta^{\mu 0}\left[(\Phi_{y+0}-\Phi_y)B_{3}^{\mu}(y) +(\Phi_y-\Phi_{y-0}) B_{3}^{\mu}(y-0)\right]\nonumber\\
&&+2\delta^{\sigma 0}\left[(\Phi_{y+0}-\Phi_y)B_{12}^{\sigma}(y)+(\Phi_y-\Phi_{y-0})B_{12}^{\sigma}(y-0)\right]\nonumber\\
&&+2\delta^{\alpha 0}\left[(\Phi_{y+0}-\Phi_y)B_{12}^{\alpha}(y)+(\Phi_y-\Phi_{y-0})B_{12}^{\alpha}(y-0)\right]\nonumber\\
&&-2(1-\delta^{\mu 0})\left(U_{\mu,y}\Phi_{y+\mu}-U_{\mu,y-\mu}^\dagger\Phi_{y-\mu}\right)B_{3}^{\mu}(y)\nonumber\\
&&+2(1-\delta^{\sigma 0})\left(U_{\sigma,y}\Phi_{y+\sigma}-U_{\sigma,y-\sigma}^\dagger\Phi_{y-\sigma}\right)B_{12}^{\sigma}(y)\nonumber\\
&&+2(1-\delta^{\alpha 0})\left(U_{\alpha,y}\Phi_{y+\alpha}-U_{\alpha,y-\alpha}^\dagger\Phi_{y-\alpha}\right)B_{12}^{\alpha}(y)\nonumber\\
&&+ 2(1-\delta^{\nu 0})(1-\delta^{\lambda 0})B_{12}^{\nu\lambda}(y)\nonumber\\&&
\times\left(U_{\nu,x}U_{\lambda,x+\nu}\Phi_{x+\nu+\lambda}+U^\dagger_{\nu,x-\nu}U^\dagger_{\lambda,x-\nu-\lambda}\Phi_{x-\nu-\lambda}-U^\dagger_{\nu,x-\nu}U_{\lambda,x-\nu}\Phi_{x-\nu+\lambda}-U_{\nu,x}U^\dagger_{\lambda,x+\nu-\lambda}\Phi_{x+\nu-\lambda}\right)\nonumber\\
&&+2\delta^{\nu 0}(1-\delta^{\lambda 0})
 \times(U_{\lambda,y+0}\Phi_{y+0+\lambda}-U_{\lambda,y}\Phi_{y+\lambda}-U^\dagger_{\lambda,y-\lambda+0}\Phi_{y-\lambda+0}+U^\dagger_{\lambda,y-\lambda}\Phi_{y-\lambda})B_{12}^{\nu\lambda}(y)\nonumber\\
&&+2\delta^{\nu 0}(1-\delta^{\lambda 0})\times(U_{\lambda,y}\Phi_{y+\lambda}-U_{\lambda,y-0}\Phi_{y-0+\lambda}-U^\dagger_{\lambda,y-\lambda}\Phi_{y-\lambda}+U^\dagger_{\lambda,y-\lambda-0}\Phi_{y-\lambda-0})B_{12}^{\nu\lambda}(y-0)\nonumber\\
&&+2(1-\delta^{\nu 0})\delta^{\lambda 0}
\times(U_{\lambda,y}\Phi_{y+0+\nu}-U_{\nu,y}\Phi_{y+\nu}-U^\dagger_{\nu,y-\nu}\Phi_{y-\nu+0}+U^\dagger_{\nu,y-\nu}\Phi_{y-\nu})B_{12}^{\nu\lambda}(y)\nonumber\\
&&+2(1-\delta^{\nu 0})\delta^{\lambda 0}\times(U_{\nu,y}\Phi_{y+\nu}-U_{\nu,y}\Phi_{y-0+\nu}-U^\dagger_{\nu,y-\nu}\Phi_{y-\nu}+U^\dagger_{\nu,y-\nu}\Phi_{y-\nu-0})B_{12}^{\nu\lambda}(y-0)\Big).\nonumber\\
\ee
We have defined
\be
c_\phi(y) &=& \left(C_\mu^0(C_\sigma^+C_\alpha^-C_\alpha^-+C_\sigma^-C_\alpha^+C_\alpha^-)D_{\nu\lambda}^+\right)_y,\\
 c_\mu(y) &=& \left(C_\sigma^+C_\alpha^-C_\alpha^-+C_\sigma^-C_\alpha^+C_\alpha^-)D_{\nu\lambda}^+\right)_y,\\
c_\sigma(y) &=& (C_\mu^0C_\alpha^-(C_\alpha^-D_{\nu\lambda}^++C_\alpha^+D_{\nu\lambda}^-))_y,\\
c_\alpha(y) &=& (C_\mu^0[2C_\sigma^-C_\alpha^+D_{\nu\lambda}^-+C_\sigma^+C_\alpha^-D^-_{\nu\lambda}+C_\sigma^- C_\alpha^- D_{\nu\lambda}^+])_y,\\
c_{\nu\lambda}(y) &=& (C_\mu^0(C_\alpha^+C_\alpha^-C_\alpha^-+C_\sigma^-C_\alpha^+C_\alpha^-))_y,\\
B_{12}^{a}(y)&=&\frac{\left(i\sigma^1\Rea(c_{a}(y))-i\sigma^2\Ima(c_{a}(y))\right)}{\phi_c^2(y)},\\
B_{3}^{a}(y)&=&\frac{\left(i\sigma^3\Rea(c_{a}(y))\right)}{\phi_c^2(y)}.
\ee


\subsection{Gauge field equation of motion\label{sec:gaugeeom}}

For the gauge fields, we use the rules
\be
\frac{\delta }{\delta A^a_{\nu,y}}U_{\mu,x}=iS^a_b\sigma^bU_{\nu,y}\delta^{xy}_{\mu\nu},\qquad \frac{\delta }{\delta A^a_{\nu,y}}U^\dagger_{\mu,x}=-iU_{\nu,y}S^a_b\sigma^b\delta^{xy}_{\mu\nu}.
\ee
The matrix $S^a_b$ is unknown, but we will not need it, as it cancels out at the end of the calculation. We assume it is invertible. We go to temporal gauge $A_0=0$ at the end. We introduce the ``electric'' field as\footnote{Some of the literature (for instance \cite{Tranberg:2006ip,Tranberg:2006dg}) states the convention
$
E_{n,x}^a=i\Tr[\sigma^a U_{n,x}U^\dagger_{n,x+0}]
$
while in fact using Eq.~(\ref{eq:efielddef}) in practice. When used consistently, the two of course lead to equivalent results.}
\be
\label{eq:efielddef}
E_{n,x}^a=-\frac{i}{2}\Tr[\sigma^a U_{n,x}U^\dagger_{n,x+0}].
\ee
We then have
\be
\label{eq:gaugeeom}
\beta_G^t\partial_0'E^a_{n,y}-\frac{\beta_G^s}{2}\sum_m D_{m}^{'ab}\Tr[i\sigma^bU_{m,y}U_{n,y+m}U_{m,y+n}^\dagger U_{n,y}^\dagger]\nonumber\\
-\beta^s_H\Tr[\left(U_{n,y}\Phi_{y+n}-\Phi_y\right)^\dagger i\sigma^a\Phi_y]
-(S_a^b)^{-1}\frac{\delta S_{\rm CP, 6}}{\delta A^a_{y,n}}=0,
\ee
where we have introduced the adjoint (backwards) covariant derivative
\be
(D^{'ab}_{m})_{xy}F^{b}_{y}=F^a_{x}-\frac{1}{2}\Tr[U_{x-m,x}\sigma^a U^\dagger_{x-m,x}\sigma^b]F^b_{x-m}.
\ee
We need to define
\be
F^b_y&=&\Phi^\dagger_yi\sigma^b U_{n,y}\Phi_{y+n},\\
H1_y^b&=&\Phi^\dagger_{y-0}i\sigma^bU_{n,y}\Phi_{y+n},\\
H2_y^b&=&\Phi^\dagger_{y+0}i\sigma^bU_{n,y}\Phi_{y+n},\\
H3_y^b&=&\Phi^\dagger_{y+n-0}U_{n,y}^\dagger i\sigma^b\Phi_{y},\\
H4_y^b&=&\Phi^\dagger_{y+n+0}U_{n,y}^\dagger i\sigma^b\Phi_{y},\\
G1_y^b&=&\Phi^\dagger_{y}i\sigma^bU_{n,y}U_{\lambda,y+n}\Phi_{y+n+\lambda},\\
G2_y^b&=&\Phi^\dagger_{y+n}U_{n,y}^\dagger i\sigma^bU_{\lambda,y-\lambda}^\dagger\Phi_{y-\lambda},\\
G3_y^b&=&\Phi^\dagger_{y+n}U_{n,y}^\dagger i\sigma^bU_{\lambda,y}\Phi_{y+\lambda},\\
G4_y^b&=&\Phi^\dagger_{y}i\sigma^bU_{n,y}U^\dagger_{\lambda,y+n-\lambda}\Phi_{y+n-\lambda},
\ee
\be
c_\mu(y)&=&D_{\nu\lambda}^+\left(C_\sigma^+C_\alpha^-C_\alpha^-+C_\sigma^-C_\alpha^+C_\alpha^-\right)_y,\\
c_\sigma(y)&=&C_\mu^0\left(D_{\nu\lambda}^+C_\alpha^-C_\alpha^-+D_{\nu\lambda}^-C_\alpha^+C_\alpha^-\right)_y,\\
c_\alpha(y)&=&C_\mu^0\left(C_\sigma^-C_\alpha^-D_{\nu\lambda}^++D_{\nu\lambda}^-(2C_\sigma^-C_\alpha^++C_\sigma^+C_\alpha^-)\right)_y,\\
c_{\nu\lambda}(y)&=&C_\mu^0\left(C_\sigma^+C_\alpha^-C_\alpha^-+C_\sigma^-C_\alpha^+C_\alpha^-\right)_y,\\
B_{12}^{a}(y)&=&\frac{\left(i\sigma^1\Rea(c_{a}(y))-i\sigma^2\Ima(c_{a}(y))\right)}{\phi_c^2(y)},\\
B_{3}^{a}(y)&=&\frac{\left(i\sigma^3\Rea(c_{a}(y))\right)}{\phi_c^2(y)}.
\ee
The last term in (\ref{eq:gaugeeom}) reads,
\be
\label{eq:Geom}
(S_a^b)^{-1}\frac{\delta S_{\rm CP, 6}}{\delta A_{n,y}^a}&=&\frac{\beta_k}{\phi^2_\Lambda(y)}\bar{\epsilon^{\mu\nu\lambda\sigma}}\big(\nonumber\\
&&-\delta^{n\mu}\big[\Tr (B_{3}^{\mu}(y)-B_{3}^{\mu}(y+n))F_y^b\big]\nonumber\\
&&+\delta^{n\sigma}\big[\Tr (B_{12}^{\sigma}(y)- B_{12}^{\sigma}(y+n))F_y^b\big]\nonumber\\
&&+\delta^{n\alpha}\big[\Tr (B_{12}^{\alpha}(y)-B_{12}^{\alpha}(y+n))F_y^b\big]\nonumber\\
&&-\delta^{n\lambda}_{0\nu}\big[\Tr (B_{12}^{\nu\lambda}(y)-B_{12}^{\nu\lambda}(y+n))(H2_y^b-H4_y^b)\big]\nonumber\\
&&-\delta^{n\lambda}_{0\nu}\big[\Tr (B_{12}^{\nu\lambda}(y+n-0)-B_{12}^{\nu\lambda}(y-0))(H1_y^b-H3_y^b)\big]\nonumber\\
&&+\delta^{n\nu}(1-\delta^{0\lambda})\big[\Tr B_{12}^{\nu\lambda}(y)(G1_y^b-G4_y^b)+\Tr B_{12}^{\nu\lambda}(y+n)(G3_y^b-G2_y^b)\nonumber\\
&&  -\delta^{n\nu}(1-\delta^{0\lambda})\big[\Tr B_{12}^{\nu\lambda}(y+\lambda)G3_y^b+\Tr B_{12}^{\nu\lambda}(y-\lambda)G2_y^b\big]\nonumber\\
&&  -\delta^{n\nu}(1-\delta^{0\lambda})\big[\Tr B_{12}^{\nu\lambda}(y+n+\lambda)G1_y^b+\Tr B_{12}^{\nu\lambda}(y+n-\lambda)G4_y^b\big]
\big).\nonumber\\
\ee


\subsection{Gauss law\label{sec:gausslaw}}
Gauss Law is the equation of motion (or rather, constraint equation) resulting from variation with respect to $A_0$. It means that the quantity
\be
G_y^a=-\beta_G^t\sum_n D_n^{'ab}E^b_{y,n}+\beta_H^t\Tr[\partial_0\Phi_y^\dagger i \sigma^a\Phi_y]-(S_a^b)^{-1}\frac{\delta S_{\rm CP, 6}}{\delta A_{0,y}^a},
\ee
is constant in time. We first define
\be
F_y^b&=&\Phi_y^\dagger  i\sigma^b\Phi_{y+0},\\
H1_y^b&=&\Phi_y^\dagger  i\sigma^b\left(U_{\lambda,y+0}\Phi_{y+0+\lambda}-U^\dagger_{\lambda,y+0-\lambda}\Phi_{y+0-\lambda}\right),\\
H2_y^b&=&\Phi_{y+0}^\dagger  i\sigma^b\left(U_{\lambda,y}\Phi_{y+\lambda}-U_{\lambda,y-\lambda}^\dagger\Phi_{y-\lambda}\right),\\
H3_y^b&=&\Phi_{y-\nu}^\dagger U_{\nu,y-\nu} i\sigma^b\Phi_{y+0}+\Phi^\dagger_{y-\nu+0}U_{\nu,y-\nu+0} i\sigma^b\Phi_{y},\\
H4_y^b&=&\Phi_{y+0+\nu}^\dagger U_{\nu,y+0}^\dagger i\sigma^b\Phi_y+\Phi^\dagger_{y+\nu}U_{\nu,y}^\dagger i\sigma^b\Phi_{y+0} ,\\
c_\mu(y)&=&\left[(C_\sigma^+C_\alpha^-+C_\sigma^-C_\alpha^+)C_\alpha^-D_{\nu\lambda}^+\right]_y,\\
c_\sigma(y)&=&\left[C_\mu^0C_\alpha^-(C_\alpha^-D_{\nu\lambda}^++C_\alpha^+D_{\nu\lambda}^-)\right]_y,\\
c_\alpha(y)&=&\left[C_\mu^0(C_\sigma^-C_0^-D_{\nu\lambda}^++C_\sigma^+C^-_0D_{\nu\lambda}^-+2C_\sigma^-C_0^+D^-_{\lambda\nu})\right]_y,\\
c_{\nu\lambda}(y)&=&\left[C_\mu^0(C_\sigma^+C_\alpha^-C_\alpha^-+C_\sigma^-C_\alpha^+C_\alpha^-)\right]_y,\\
B_{12}^{a}(y)&=&\frac{\left(i\sigma^1\Rea(c_{a}(y))-i\sigma^2\Ima(c_{a}(y))\right)}{\phi_c^2(y)},\\
B_{3}^{a}(y)&=&\frac{\left(i\sigma^3\Rea(c_{a}(y))\right)}{\phi_c^2(y)},
\ee
and then we have
\be
\label{eq:Gauss}
(S_a^b)^{-1}\frac{\delta S_{\rm CP, 6}}{\delta A_{0,y}^a}&=&\frac{\beta_\kappa}{\phi^2_\Lambda(y)}\bar{\epsilon^{\mu\nu\lambda\sigma}}\Bigg(
-2\delta^{\mu 0} \Tr F_y^bB_{3}^{\mu}(y)
+2\delta^{\sigma 0}\Tr F_y^bB_{12}^{\sigma}(y)
+2\delta^{\alpha 0}\Tr F_y^bB_{12}^{\alpha}(y)\nonumber\\&&
\delta^{\nu 0}\Tr (H1_y^b+H2_y^b)B_{12}^{\nu\lambda}(y)-\delta^{\lambda 0}\Tr H4_y^bB_{12}^{\nu\lambda}(y+\nu)+\delta^{\lambda 0}\Tr H3_y^bB_{12}^{\nu\lambda}(y-\nu)\Bigg).\nonumber\\
\ee


\section{Numerical procedure\label{sec:procedure}}

The equations of motion are implicit in time, but because of the specific choice of time-discretization, there are only nearest-neighbour couplings in time, while there are next-to-nearest neighbour couplings in space. This makes it possible to solve the equations of motion by iteration in the CP-violating force term. Because the size of the CP-violating term is small, this converges under certain conditions on $\kappa^{\rm CP}$ and $dt$. 

We use a lattice of $V=L^3$ sites and a lattice spacing $a_x$, with
\be
a_xm_H=0.35,\quad m_H/m_W=2, \quad L=64, \quad g^2=4/9, \quad dt=a_t/a_x=0.0125, \nonumber\\
\ee
and we run the simulations until the system settles, $m_Ht_{\rm stop}=30$. We will choose $\kappa_{CP}$ to get an observable signal at a given cut-off, and then scale back to a common $\kappa^{\rm CP}$ value assuming a linear dependence\footnote{A full-fledged determination of the $\kappa^{\rm CP}$ dependence is currently beyond our reach, numerically.}. The combination of large lattices, small timestep and the need to iterate makes this a numerically very heavy problem, requiring $\mathcal{O}(10^6)$ CPU hours on state-of-the art computer clusters.


\subsection{Initial conditions\label{sec:initcond}}

We generate an ensemble of initial conditions, with vanishing gauge field $A_\mu=0$, and the Higgs field reproducing the quantum vacuum in the pre-quench potential $V(\phi)=+\mu^2\phi^{\dagger}\phi$, 
\be
\langle\phi_k^{\dagger}\phi_k\rangle_{t=0}=\frac{1}{2\sqrt{k^2+\mu^2}},\quad \langle\partial_t\phi_k^{\dagger}\partial_t\phi_k\rangle_{t=0}=\frac{\sqrt{k^2+\mu^2}}{2}, \quad|k|<|\mu|.
\ee
Only the unstable Higgs modes are initialised, in order to mimic that the quantum modes grow due to the spinodal instability, and that we only keep the modes that thus become classically large \cite{Rajantie:2000nj,GarciaBellido:2002aj,Smit:2002yg}. The spinodal instability affects all the modes with $k^2_{\rm lat}<\lambda v^2$, which means sets of integers $k_x$, $k_y$, $k_z$, satisfying
\be
\frac{(La_xm_H)^2}{8\pi^2}>k_x^2+k_y^2+k_z^2.
\ee
For our choice, we have $\sim 80$ unstable modes, enough that the IR dynamics is well represented.

The initialisation is generated by Monte-Carlo sampling in order that the total charge on the lattice is zero. However, the local charge density is non-zero, and the gauge momentum, the electric field $E_{n}^a$, follows from the Gauss constraint, given the Higgs background,
\be
\partial_n'E_n^a=i\frac{2\beta_H^t}{\beta_G^t}\Tr[(\Phi_{x+0}-\Phi_x)^\dagger\sigma^a\Phi_x].
\ee
The solution to the Gauss constraint makes the approximation that $\kappa_{CP}=0$. Whereas in \cite{Tranberg:2003gi,Tranberg:2006ip,Tranberg:2006dg} the CP-violation vanishes when $A_\mu=0$, in the present case, there is a nonzero term which includes the electric field $E$, namely
\be
\label{eq:initGauss}
(S_a^b)^{-1}\frac{\delta S_{\rm CP, 6}}{\delta A_{0,y}^a}_{A_\mu=0}&=&\frac{\beta_\kappa}{\phi^2_\Lambda(y)}\bar{\epsilon^{\mu\nu\lambda\sigma}}\Bigg(
\delta^{\nu 0}\Tr (H1_y^b+H2_y^b)B_{12}^{\nu\lambda}(y)\nonumber\\&&-\delta^{\lambda 0}\Tr H4_y^bB_{12}^{\nu\lambda}(y+\nu)+\delta^{\lambda 0}\Tr H3_y^bB_{12}^{\nu\lambda}(y-\nu)\Bigg).\nonumber\\
\ee
Solving the full Gauss constraint now becomes a non-linear minimization process, which we do not attempt here. We note that we do not observe any effect of this approximation. The asymmetry generated does not arise in the early stages of the simulation, but after the initial Higgs roll-off. In the absence of CP-violation, the Chern-Simons number averaged over the initial ensemble is indeed zero.


\subsection{Observables\label{sec:observables}}

We will monitor the Chern-Simons number\footnote{The plaquettes are symmetrized forwards and backwards in space and time as for the action.},
\be
\label{eq:anomaly}
N_{\rm CS}(t)-N_{\rm CS}(0)=\frac{1}{16\pi^2}\sum_{0}^t\sum_x\epsilon^{\mu\nu\rho\sigma}\Tr\left[U_{\mu\nu,x}U_{\rho\sigma,x}\right],
\ee
the Higgs winding number,
\be
N_{\rm w}=\frac{1}{192\pi^2 V}\sum_{x,ijk}\epsilon_{ijk}\Tr\left[\left(M_{x+i}-M_{x-i}\right)M^\dagger_{x}\left(M_{x+j}-M_{x-j}\right)M^\dagger_{x}\left(M_{x+k}-M_{x-k}\right)M^\dagger_{x}\right],\nonumber\\
\ee
with $M=\Phi/|\phi|$, and the average Higgs field,
\be
\label{eq:Higgs}
\bar{\phi^2}=\frac{1}{V}\sum_{x}\frac{1}{2}\Tr\,\Phi_x^\dagger\Phi_x.
\ee
Through the anomaly equation, the Chern-Simons number (\ref{eq:anomaly}) is directly related to the baryon number of fermions living in the gauge field background. However, in the vacua the Chern-Simons number is integer and equal to the Higgs winding number. The latter is a much cleaner observable since it is always integer\footnote{Up to lattice errors of order 10 percent, and up to transients related to integer ``flips''.} and settles early on. We therefore use that at asymptotically late times
\be
B(t)-B(0)=3\left(N_{\rm cs}(t)-N_{\rm cs}(0)\right)=3\left(N_{\rm w}(t)-N_{\rm w}(0)\right).
\ee
In addition, in order to minimize the statistical noise, we average over a strictly CP-symmetric ensemble of initial field configuration, including for each random initial configuration its C(P) conjugate. In practice, this amounts to running each configuration first with $+\beta_\kappa$, calculating the Chern-Simons $N_{\rm cs}$ and the Higgs winding number $N_{\rm w}$. And then running again with $-\beta_\kappa$, calculating $-N_{\rm cs}$ and $-N_{\rm w}$. In this way we can quantify the CP-asymmetry configuration by configuration, by calculating the integer
\be
\Delta N_{\rm w}=N_{\rm w}(\beta_\kappa)-N_{\rm w}(-\beta_\kappa),
\ee
and perform the statistics on these, rather than the $N_{\rm w}$ individually. This has the advantage of reducing the statistical errors, since we note that in practice $\Delta N_{\rm w}$ is either 0, $\pm 1$ or $\pm 2$. A more detailed exposition of this procedure can be found in \cite{Tranberg:2006ip}.


\section{Results\label{sec:results}}


\subsection{Size of the force\label{sec:forceterm}}

To gather insight into the size of the CP-violating effect and the impact of the cut-off, it is instructive to first perform simulations using CP-{\it conserving} dynamics, while at the same time calculating the CP-violating force from the gauge equation of motion. Writing the discretized equation (\ref{eq:Geom}) as
\be
E_\mu^a(x,t) = E_\mu^a(x,t-dt)+\delta E_\mu^{a,0}(x,t)+\beta_\kappa\delta E_\mu^{a,1}(x,t).
\ee
in terms of a CP-symmetric $\delta E_\mu^{a,0}$ and a CP-breaking force component $\beta_\kappa\delta E_\mu^{a,1}$ we average the latter\footnote{We checked that the same picture emerges when adding up in quadrature or absolute values of the force.} 
\be
\frac{1}{V}\sum_x \beta_\kappa\delta E_\mu^{a,1}(x,t),
\ee
for all values of $a=1,2,3$, $\mu=1,2,3$.
We vary the cut-off $\sqrt{c}\Lambda$ through the values
\be
\sqrt{c}\Lambda=\{0,~1,~3,~10,~30,~90\}\, \textrm{GeV}.
\ee
\begin{figure}
\begin{center}
\epsfig{file=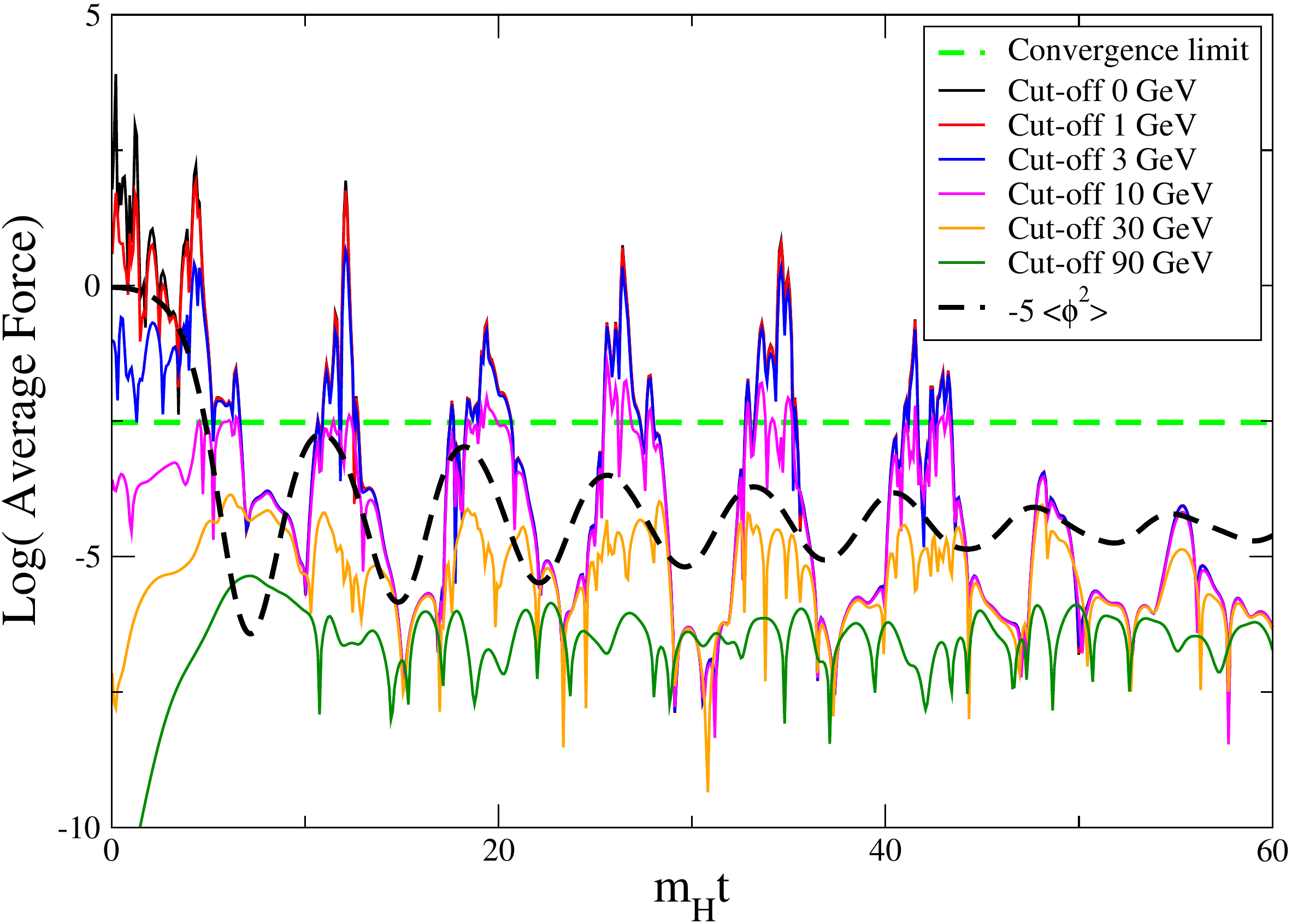,width=11cm,clip}
\caption{The logarithm of the average force term in a run with $dt=0.05$, $\kappa=10$, cut-offs 0 (black), 1, 3, 10, 30, 90 (green) GeV.  Superposed, the average Higgs field; to fit into the picture, multiplied by -5 (black dashed).\label{cutoffdep}}
\end{center}
\end{figure}

Fig.~\ref{cutoffdep} shows the average force in time. The black dashed line is (for clarity, a rescaled and inverted version of) the Higgs field squared, averaged over the lattice (\ref{eq:Higgs}). $\phi^\dagger\phi$ ``rolls off'' the potential hill from zero to some finite value, around which it oscillates. This value is the broken phase expectation value at finite temperature, in this normalisation about -4.9. For comparison, the zero temperature vev would correspond to -5.

The oscillations of the averaged Higgs length disguises that each field configuration is highly inhomogeneous. As was shown in \cite{GarciaBellido:2003wd,vanderMeulen:2005sp}, there are many zeros initially and a few whenever the average Higgs amplitude is small. The distribution of zeros is in turn strongly dependent on the value of the Higgs mass \cite{vanderMeulen:2005sp}. 

At such nuclei, where the Higgs field length is very small, we expect an amplification of the force term. Indeed we see in Fig.~\ref{cutoffdep} that the force is largest in peaks corresponding to the extrema of Higgs oscillations, with a small delay compared to the actual minima. This feature corresponds to the observation that the maximum number of Higgs zeros also lags behind the minima of the oscillations of the averaged Higgs length \cite{vanderMeulen:2005sp}.

Without a cut-off (black) the averaged force is very large, especially initially. By studying the distribution of the force in space, we found that the average is indeed dominated by a few isolated points. The force is also strongly peaked in time, following the Higgs oscillations.
As we introduce a cut-off and increase it, as long as $\sqrt{c}\Lambda<10\,$GeV, we see no major difference (red, blue lines). But for larger cut-off ($10-30\,$GeV, magenta, orange lines), the peaks are cut down. Finally at a cut-off of $90\,$GeV, the peaks are almost completely gone. 

On a technical note, the light green dashed line is a representation of the value above which the iteration algorithm has been seen to not converge.  At the values of $\kappa^{\rm CP}$ used, we had 4-6 iterations per timestep.

The force is strictly proportional to the CP violation prefactor, $\kappa^{\rm CP}$. To get maximum asymmetry for a given cut-off, we should use a $\kappa^{\rm CP}$ bringing the force close to the dashed green line. On a lattice of the size used here, a few field configurations will then produce an asymmetry, and we should interpolate back to the "physical" value of $\kappa^{\rm CP}=9.87$ to find the actual baryon asymmetry. We will do so assuming that the dependence of the asymmetry on $\kappa^{\rm CP}$ is linear. We note that the peak values of the averaged force roughly scales as $\Lambda^{-4}$. Still, this is an averaged quantity, and it may  be that at a nucleus the scaling is stronger. 


\subsection{Single trajectories\label{sec:singletraj}}

\begin{figure}
\begin{center}
\epsfig{file=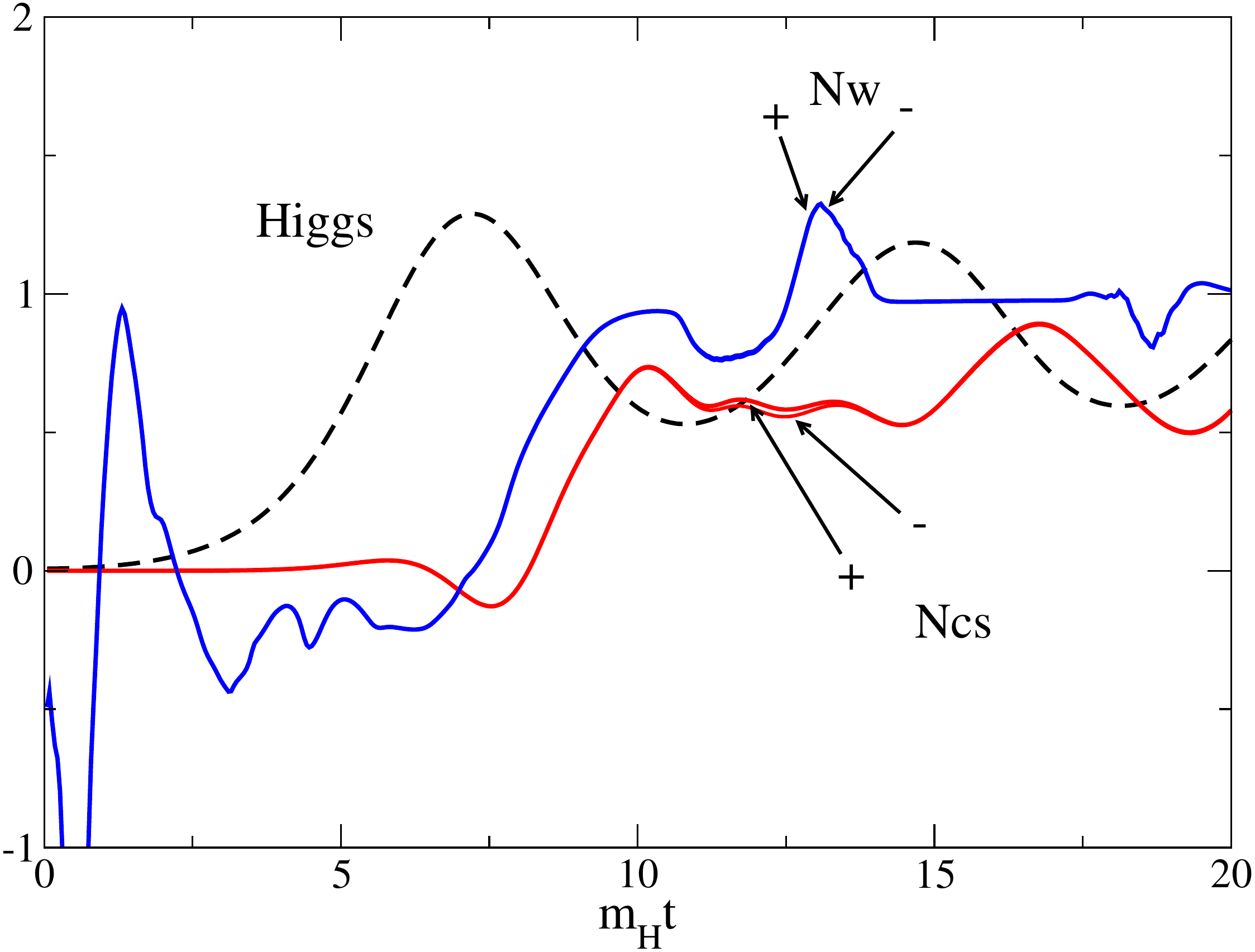,width=7cm,clip}
\epsfig{file=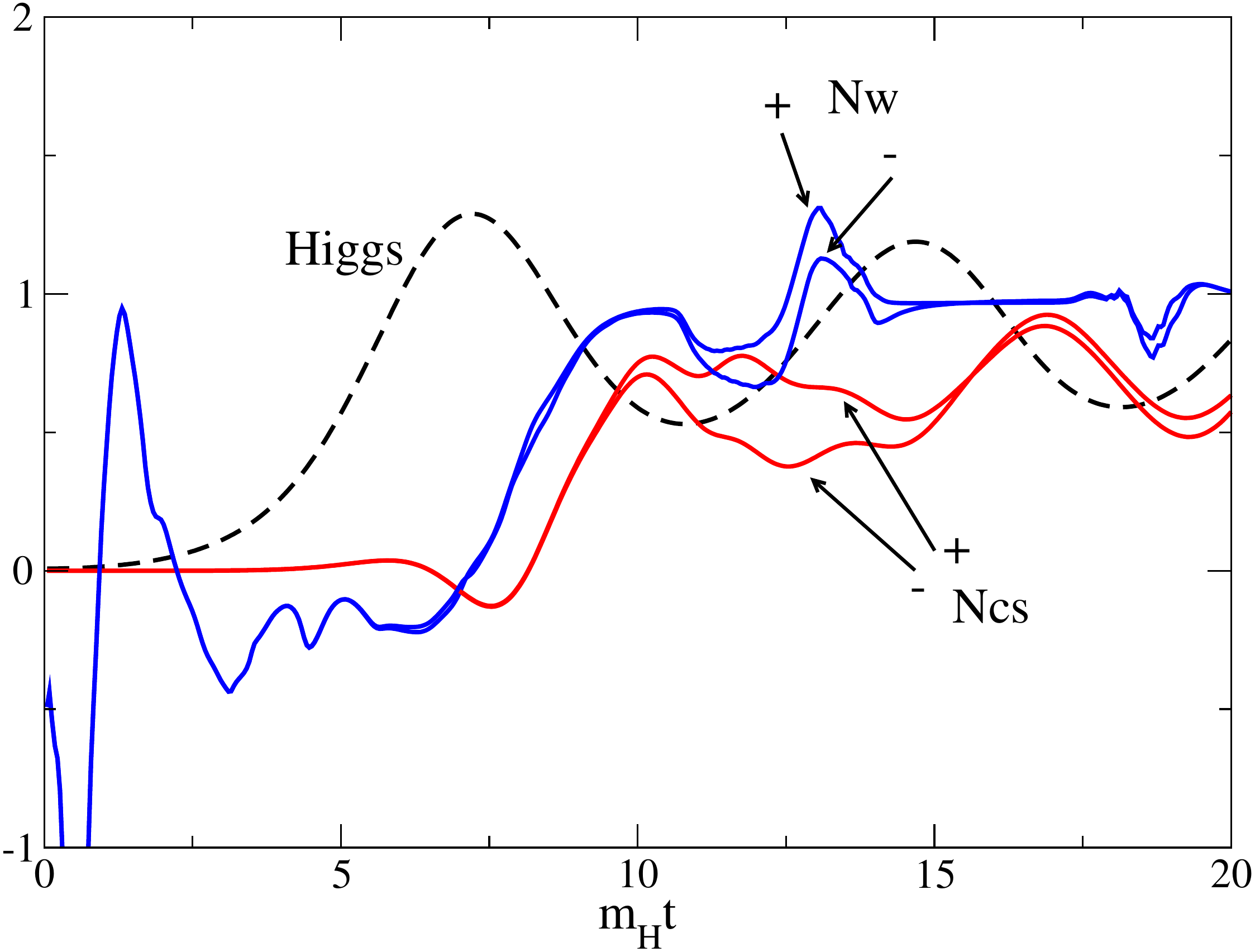,width=7cm,clip}
\caption{Left: The average Higgs length (black), the Chern-Simons number (red) and the Higgs winding number (blue) for a pair of trajectories with $\kappa^{\rm CP}=\pm 50$. The cut-off is $\sqrt{c}\Lambda=100\,$GeV. Right: The average Higgs length (black), the Chern-Simons number (red) and the Higgs winding number (blue) for the same pair of trajectories as in on the left with $\kappa^{\rm CP}=\pm 50$. The cut-off is now $\sqrt{c}\Lambda=50\,$GeV\label{cutdep_100}}
\end{center}
\end{figure}

Let us consider the effect of CP-violation on a single pair of trajectories, an example of which is shown in Fig.~\ref{cutdep_100} (right). The cut-off is $\sqrt{c}\Lambda=100\,$GeV, and we show the Higgs length (black, dashed line), the Chern-Simons number $N_{\rm cs}$ (red) and the Higgs winding number $N_{\rm w}$ (blue). Shown are two trajectories, corresponding to $\kappa^{\rm CP}=\pm 50$. The Higgs field falls off the potential hill, and starts oscillating around the minimum. Meanwhile the Chern-Simons number and winding number of the two configurations move closely together, except for a small wobble in the Chern-Simons number just after the first minimum of the Higgs oscillation. As we have seen in the previous section, this coincides with the occurrence of Higgs zeros and a peak in the CP-violating force. 

In Fig.~\ref{cutdep_100} (right) we show the exact same configuration, but run with cut-off $\sqrt{c}\,\Lambda=50\,$GeV. Now the discrepancy between the two sets of lines is much bigger and present in both the Chern-Simons number and the Higgs winding. In this case, the CP-violating force is not large enough to drive the Higgs winding to different final integers, and so $\Delta N_{\rm w}=0$.

\begin{figure}
\begin{center}
\epsfig{file=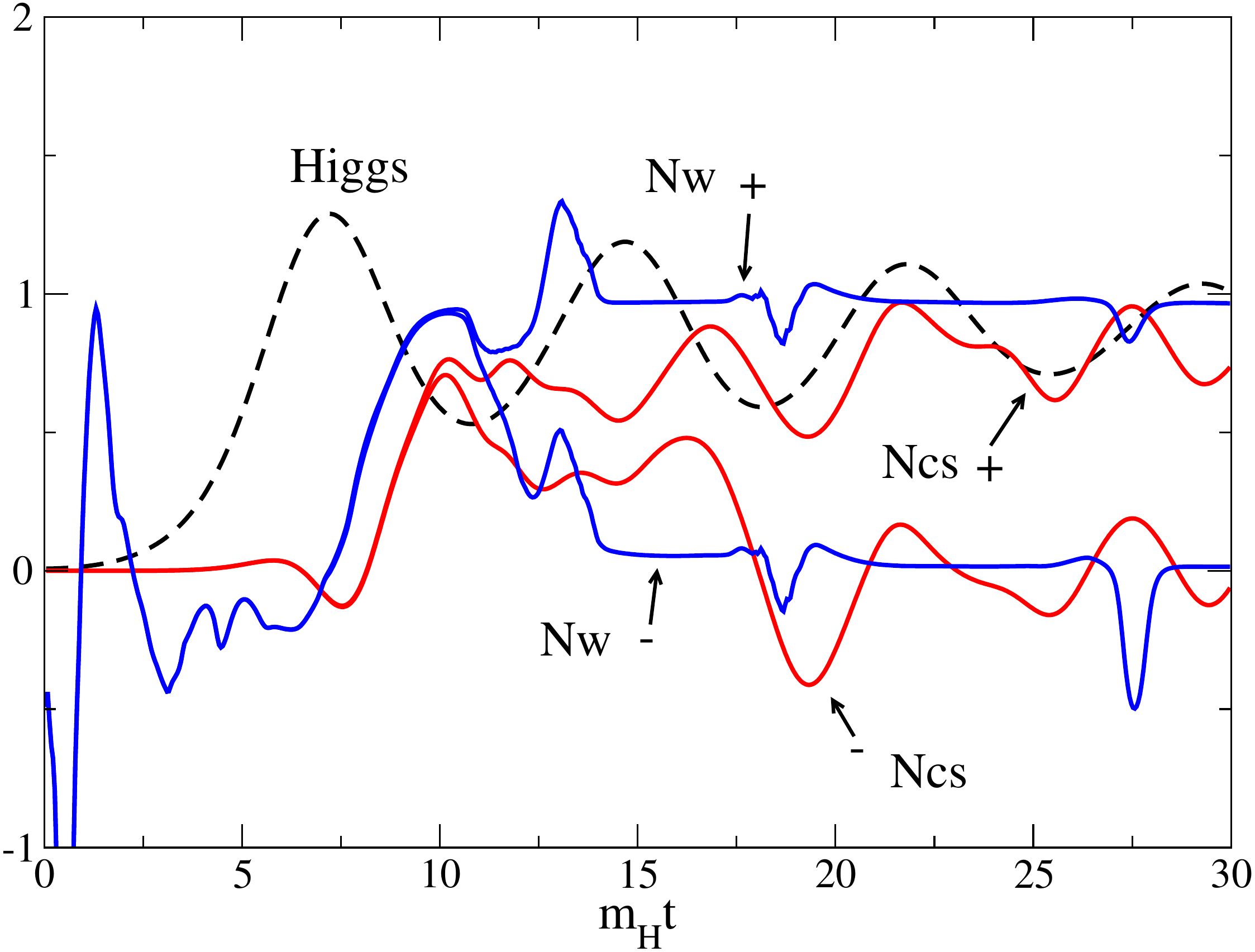,width=8cm,clip}
\caption{The average Higgs length (black), the Chern-Simons number (red) and the Higgs winding number (blue) for the same pair of trajectories as in Fig. 3 and 4, with $\kappa^{\rm CP}=\pm 800$. The cut-off is $\sqrt{c}\Lambda=100\,$GeV. We have an asymmetry, $\Delta N_{\rm w}=+1$.
\label{single1}}
\end{center}
\end{figure}

In Fig.~\ref{single1}, we show again the same pair of trajectories but now for $\kappa^{\rm CP}=800$, cut-off $\sqrt{c}\,\Lambda=100\,$GeV. Indeed, we do now have an asymmetry, $\Delta N_{\rm w}=+1$. We also see that 
 the Chern-Simons number is smooth, as a result of there being very little UV noise. All the power is in the IR tachyonic modes, until times one or two orders of magnitude longer than considered here \cite{GarciaBellido:2003wd,Skullerud:2003ki,DiazGil:2007dy}. There is therefore no need to ``cool'' the gauge field in order to obtain a reliable value for the Chern-Simons number, in contrast to simulations of the (classical) equilibrium sphaleron rate (see for instance \cite{Bodeker:1999gx}). The Higgs winding number is expected to be discrete up to lattice discretization artefacts. This is indeed the case at late times, and we consider the transition over and done by time $m_Ht=30$. Most of the action seems to take place at the first minimum in the Higgs oscillation (see also below) around $m_Ht=10$. At early times, the Higgs field is full of zeros and the Higgs winding is ill-defined, hence the wild behaviour up to the point when the Higgs field has a achieved a significant fraction of its broken phase value.
Finally, the Chern-Simons number and Higgs winding appear to roughly move together and at the same time. The Chern-Simons number is not constrained to be integer, except in the vacua. Also only dynamical constraints force it to follow the Higgs winding, thus minimizing the covariant derivative term in the energy. This also makes the Higgs winding number the cleaner observable.


\subsection{Ensemble averages\label{sec:averages}}

\begin{figure}
\begin{center}
\epsfig{file=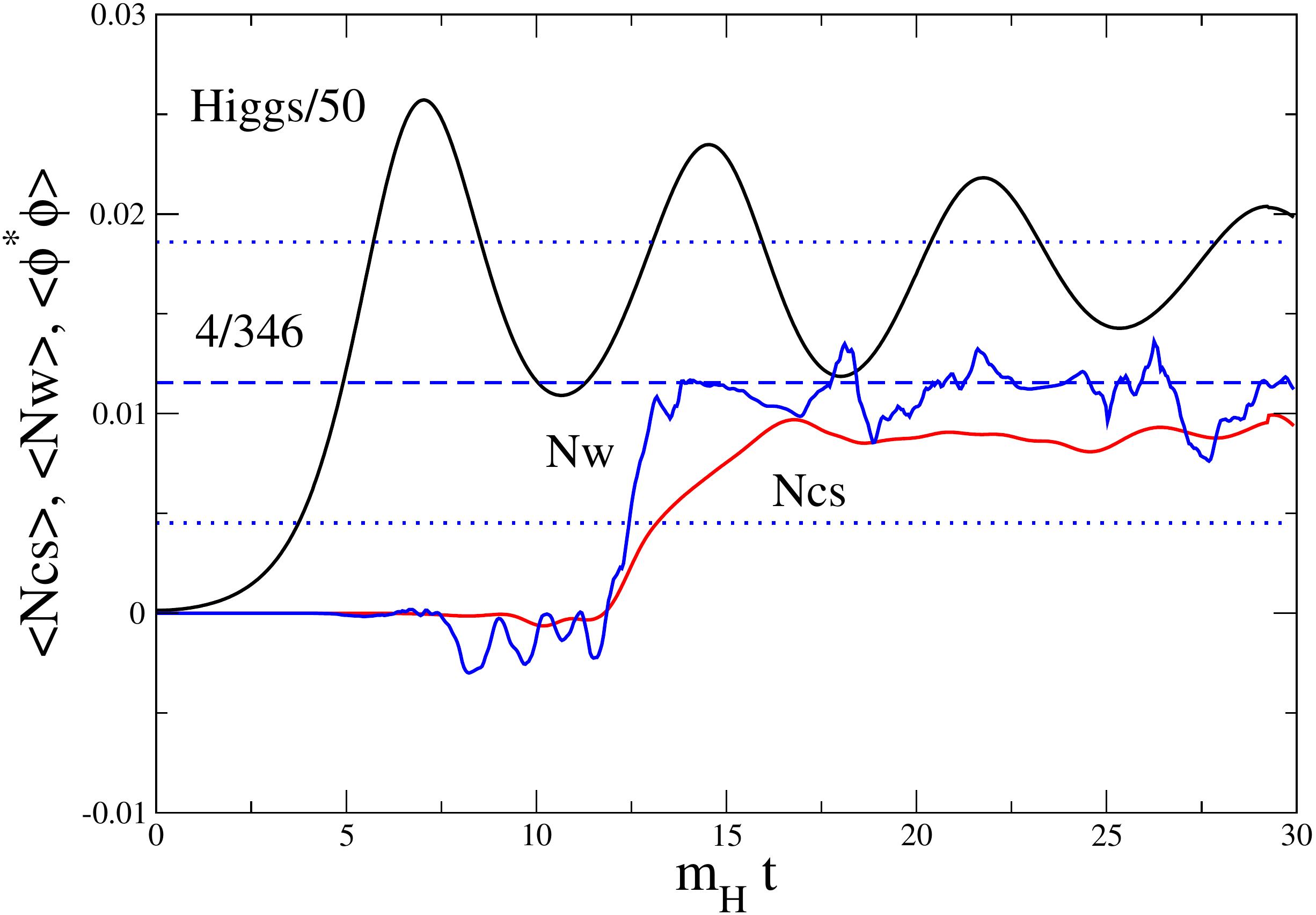,width=7cm,clip}
\epsfig{file=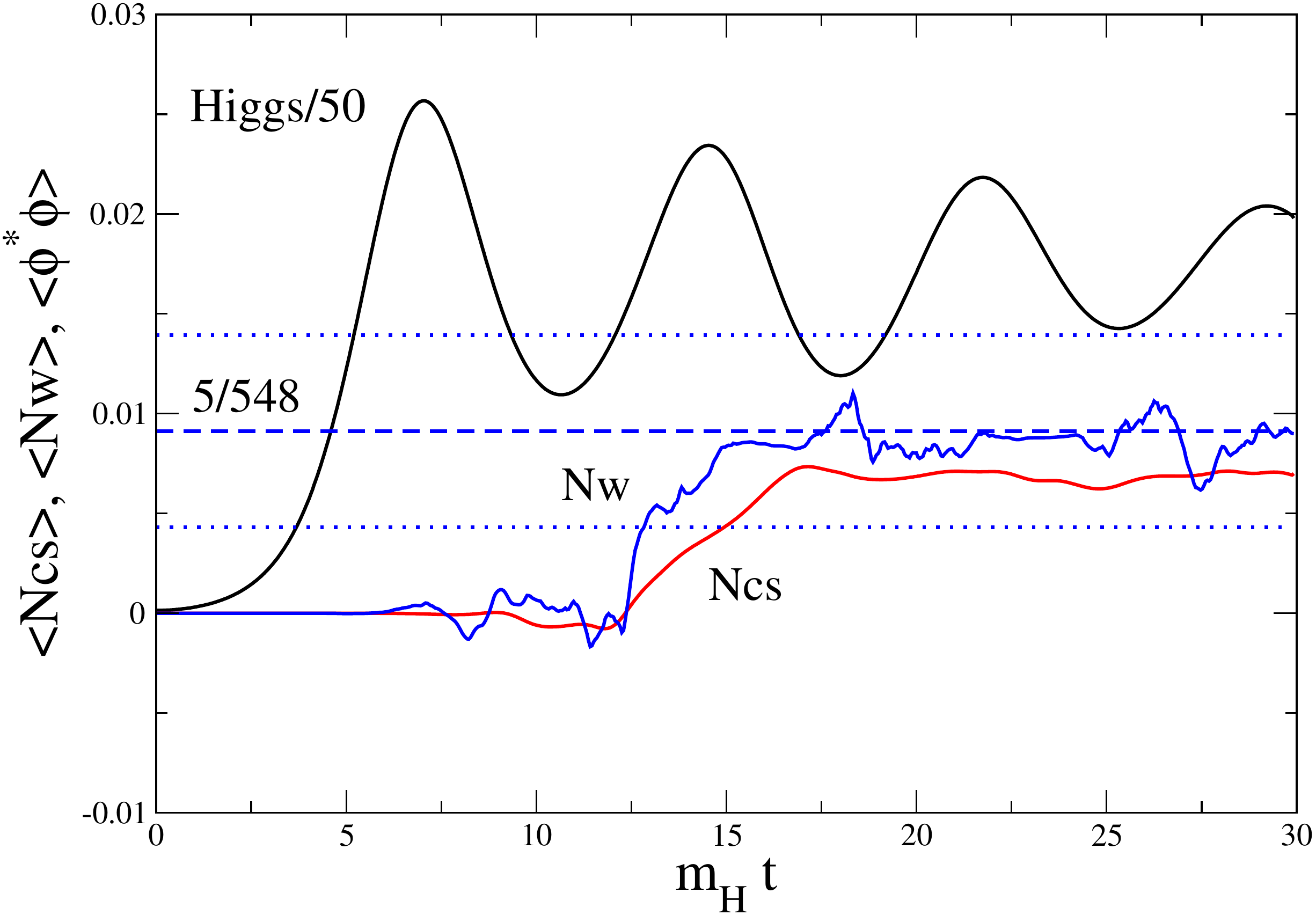,width=7cm,clip}
\caption{Left: Ensemble averages of the Higgs length (black), the Chern-Simons number (red) and the Higgs winding number (blue) with $\kappa^{\rm CP}=\pm 250$ and the cut-off is $\sqrt{c}\Lambda=75\,$GeV. Right: Ensemble averages of the Higgs length (black), the Chern-Simons number (red) and the Higgs winding number (blue) with $\kappa^{\rm CP}=\pm 800$ and the cut-off is $\sqrt{c}\Lambda=100\,$GeV.\label{average100}}
\end{center}
\end{figure}

We perform simulations of an ensemble of (pairs of) initial configurations, and determine from each pair the integer $\Delta N_{\rm w}$, which we then average over. Fig.~\ref{average100} show the average Higgs field (full black), the Chern-Simons number (red) and the Higgs winding number (full blue) for a cut-off of $75\,$GeV and $100\,$GeV. The simulations were done with $\kappa^{\rm CP}=250$ and $800$, respectively. $N_{\rm cs}$ and $N_{\rm w}$ seem to move at the same time, and also in the ensemble averaged observables there is a clear asymmetry, largely created during the first minimum of the Higgs oscillations. This is due to the appearance of actual Higgs zeros, where winding happens easily, but may also be because the CP-violation is larger in the presence of such points. In fact, since a factor of 100/75 in $\sqrt{c}\Lambda$ requires a factor of 800/250 in $\kappa^{\rm CP}$ to get roughly the same asymmetry, the result is certainly sensitive to the cut-off. More about this in the next section. 

The dashed blue line is the average of $\Delta N_{\rm w}$, with the dotted lines the error bars on that number (standard deviation). The results are presented in Table \ref{tab:results}.
\begin{table}
\begin{center}
\label{tab:results}
\begin{tabular}{|c|c|c|c|c|c|c|}
\hline
$\sqrt{c}\Lambda$&$\kappa_{\rm sim}^{\rm CP}$&$+1$&$-1$&$N_{\rm tot}$&$\frac{n_B}{n_\gamma}(\kappa^{\rm CP}=9.87)$&$\kappa^{\rm CP}_{\rm obs}$\\
\hline
$50\,\textrm{GeV}$&$50$&5&1&247&$2.4\times 10^{-6}$&$2.4\times 10^{-3}$\\
\hline
$75\,\textrm{GeV}$&$250$&5&1&173&$6.9\times 10^{-7}$&$8.5\times 10^{-3}$\\
\hline
$100\,\textrm{GeV}$&$800$&6&1&274&$1.7\times 10^{-7}$&$3.5\times 10^{-2}$\\
\hline
$125\,\textrm{GeV}$&$2800$&5&2&288&$3.4\times 10^{-8}$&$0.18$\\
\hline
$174\,\textrm{GeV}$&-&-&-&-&$4.9\times 10^{-9}$&$1.2$\\
\hline
\end{tabular}
\caption{The cut-off, $\kappa^{\rm CP}$ in the simulation, the number of $\Delta N_{\rm w}=\pm 1$ configurations, total number of configurations $N_{\rm tot}$, the corresponding asymmetry at $\kappa^{\rm CP}=9.87$, and the required $\kappa^{\rm CP}$ to match observations, all assuming a linear dependence on $\kappa^{\rm CP}$. The last line is an extrapolation to $\sqrt{c}\,\Lambda=246/\sqrt{2}\,\textrm{GeV}$ assuming an exponential depencence on the cut-off.}
\end{center}
\end{table}


\subsection{Cut-off dependence\label{sec:cutoffdep}}

\begin{figure}\begin{center}
\epsfig{file=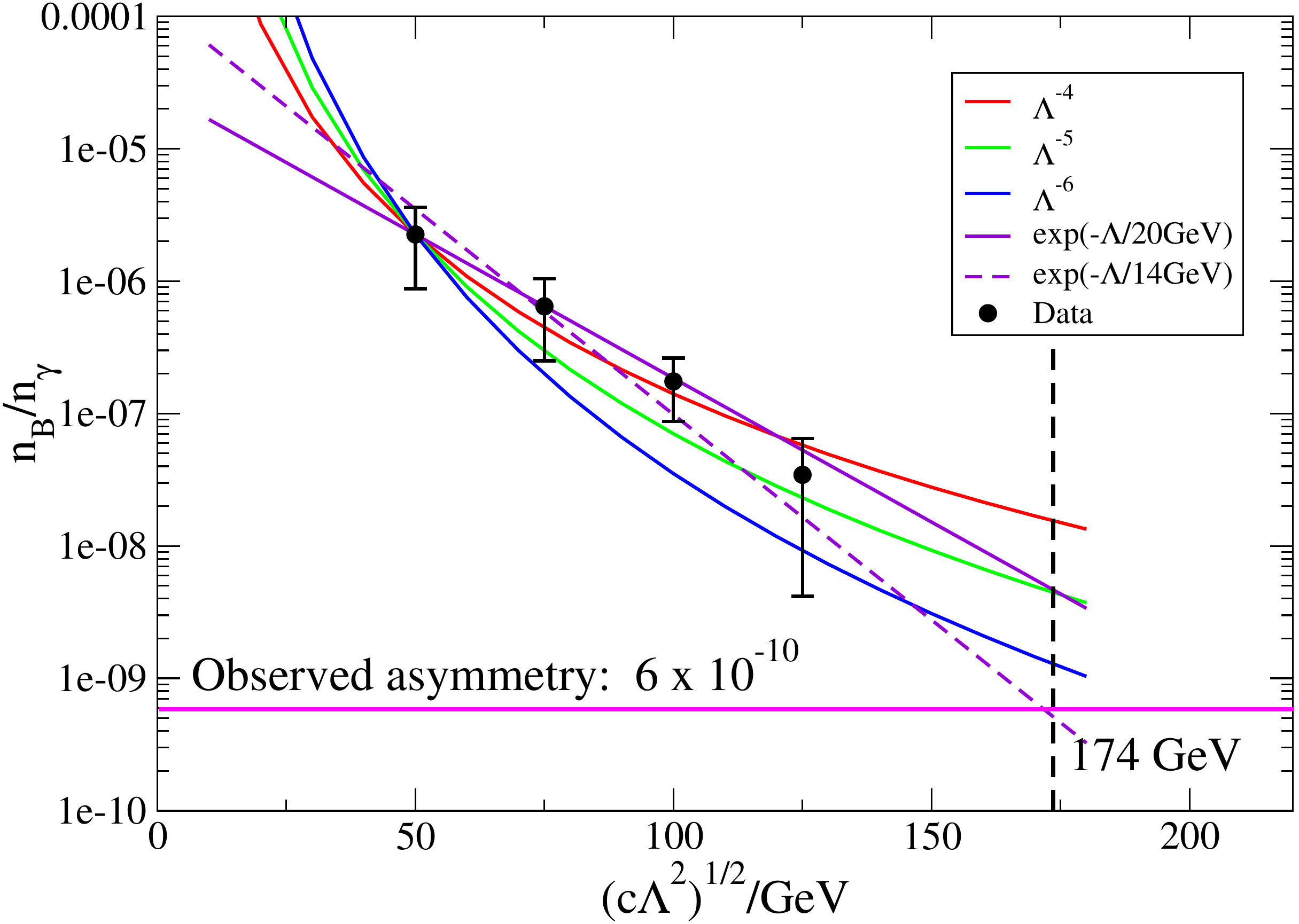,width=11cm,clip}
\caption{The dependence of the final baryon asymmetry $n_B/n_\gamma$ on the cut-off $\sqrt{c}\,\Lambda$. The purple lines are exponential forms (note the log-scale), while the red, green and blue lines are power laws. The horizontal pink line is the observed asymmetry, the vertical black-dashed line the maximum cut-off $v/\sqrt{2}$.  \label{fig:cutoffdep_full}}
\end{center}
\end{figure}

We can convert the average the Chern-Simons number to a baryon asymmetry, by writing
\be
\langle n_B\rangle=\frac{3\langle \Delta N_{\rm cs}\rangle}{L^3} = \frac{3\langle \Delta N_{\rm w}\rangle}{L^3}.
\ee
The density of photons is given as
\be
n_\gamma=\frac{1}{7.04}\frac{2\pi^2}{45} g^*T^3_{\rm reh},
\ee
in terms of the number of relativistic degrees of freedom $g^*=86.25$ and the reheating temperature $T_{\rm reh}$. This is in turn given by distributing the Higgs potential energy on the degrees of freedom,
\be
V_0=\frac{\lambda v^4}{4}=\frac{\pi^2}{30}g^*T_{\rm reh}^4.
\ee

Fig.~\ref{fig:cutoffdep_full} is the main result of this paper, and shows the dependence of the final baryon asymmetry as a function of the cut-off $\sqrt{c}\Lambda$, rescaled to $\kappa^{\rm CP}=9.87$.  There is a fairly strong dependence, best approximated by an exponential $n_B\propto \exp(\sqrt{c}\Lambda/20\,\textrm{GeV})$, but also consistent with a power law $n_B\propto (\sqrt{c}\,\Lambda)^{-4,-5,-6}$. 

Clearly, without knowledge of the actual effective cut-off, we cannot make a definitive prediction about the precise baryon asymmetry. However, for the entire allowed range of cut-offs (left of the vertical dashed line), the asymmetry rescaled to $\kappa^{\rm CP}=9.87$ is at least an order of magnitude larger than the observed value of \cite{WMAP}
\be
\frac{n_B}{n_\gamma}=6\times 10^{-10},
\ee
represented here by the pink horizontal line. An alternative way of stating the result is that for any cut-off, there is a $\kappa^{\rm CP}$ for which the asymmetry is equal to the observed one; and this  $\kappa^{\rm CP}$ is always less than the zero-temperature value of $9.87$. See also Table \ref{tab:results}.


\section{Conclusion\label{sec:conclusion}}

We have performed complete non-perturbative simulations of a cold, fast electroweak transition in the presence of Standard Model CP-violation, represented by (\ref{eq:CP_8}). A non-physical effect of Higgs zeros is dealt with by a cut-off $\Lambda$, and we calculate the resulting baryon asymmetry and its dependence on $\Lambda$. We again stress that this is {\it not} a momentum cut-off (see section \ref{sec:cut-off}). When taking the coefficient $\kappa^{\rm CP}$ to have the zero-temperature value reported in \cite{Hernandez:2008db}, we find an asymmetry larger than the observed one by a factor of between $10$ and $10^4$, depending on the cut-off. 

The total computer time used for the results presented here was approximately $10^6$ single-CPU hours on a state-of-the-art computer cluster. Although it may be possible to speed up the code somewhat, this makes large-scale parameter sweeps in Higgs mass, quench time and $\kappa^{\rm CP}$ prohibitively computer intensive. The lattice size was chosen so that each field configuration could fit on a single CPU, making parallelisation unnecessary. We believe that the lattice is big enough to correctly reproduce the IR dynamics (the spinodal instability), in which case a scaling of lattice volume is equivalent to the same scaling of the number of configurations\footnote{See for instance \cite{Smit:2002yg} for a simple 1+1 dimensional test.}.

There remains several avenues for improvement: Firstly, the cut-off introduced shields us from the breakdown of the gradient expansion near Higgs zeros. It would be interesting to attempt a calculation of this effective cut-off, or even better find a suitable (non-gradient) expansion near such nuclei. It would be interesting to see whether the same CP-violating term also appears at leading order in such an expansion. 
Secondly, so far the coefficient $\kappa^{\rm CP}$ has only been calculated at zero temperature, i.e. with zero external momenta running into the fermion loop. Although the initial condition before the electroweak transition is cold, during the spinodal roll-off, Higgs and gauge fields are far from equilibrium but also far from zero temperature. Hence, it would be appropriate to calculate $\kappa^{\rm CP}$ in the background of ``post-spinodal'' gauge-Higgs fluctuations. This is likely to be smaller than at zero temperature, but how much depends on the details of the transition, and in particular the speed of the quench. 

Ultimately, one would like to include the fermion fields themselves in the dynamics, with the full CKM matrix. This implies  implementing fermions on the lattice in real-time; on the other hand there will then be no problems with Higgs zeros or numerically complicated bosonic terms.

In simulations with the leading order CP-violating term\footnote{Which is absent in the Standard model}, sufficient baryon asymmetry implies a bound on the quench rate \cite{Tranberg:2006dg},
\be
v=\frac{1}{2\mu^3}\frac{d\mu_{\rm eff}^2(t)}{dt}>0.1.
\ee
A similar requirement will apply to the sixth order term considered here, providing another natural extension of the present work. 

In existing scenarios of Cold Electroweak Baryogenesis, the Higgs transition is triggered by a second scalar field. Although this would introduce model-dependence, including the dynamics of this additional scalar field in the simulations is also an obvious next step (see also \cite{GarciaBellido:2003wd,DiazGil:2007dy} for detailed simulations of such a system, but without CP-violation). 

For many years, Standard model CP-violation has been ruled out as a source for baryogenesis. However, if baryogenesis was cold enough, so that the zero-temperature calculation of $\kappa^{\rm CP}$ can be trusted (or the similar calculation for CP-violating terms in the CP-even sector), it seems that the observed baryon asymmetry may in fact be accounted for without the addition of new sources of CP-violation.

\appendix

\begin{acknowledgments}
I am indebted to Michael G. Schmidt, Andres Hernandez and Thomas Konstandin for pleasant and fruitful collaboration and discussion. I thank Misha Shaposhnikov and Kari Rummukainen for useful discussions, and acknowledge support from Academy of Finland Grants 114371 and 1134018. The numerical work was performed on the Murska cluster of CSC, the Finnish supercomputing center.
\end{acknowledgments}



\begin{thebibliography}{99}

\bibitem{Hernandez:2008db}
  A.~Hernandez, T.~Konstandin and M.~G.~Schmidt,
  Nucl.\ Phys.\  B {\bf 812} (2009) 290
  [arXiv:0810.4092 [hep-ph]].

\bibitem{Tranberg:2009de}
  A.~Tranberg, A.~Hernandez, T.~Konstandin and M.~G.~Schmidt,
  Phys.\ Lett.\  B {\bf 690} (2010) 207
  [arXiv:0909.4199 [hep-ph]].


\bibitem{Kuzmin:1985mm}
  V.~A.~Kuzmin, V.~A.~Rubakov and M.~E.~Shaposhnikov,
  Phys.\ Lett.\  B {\bf 155} (1985) 36.

\bibitem{Rubakov:1996vz}
  V.~A.~Rubakov and M.~E.~Shaposhnikov,
  Usp.\ Fiz.\ Nauk {\bf 166} (1996) 493
  [Phys.\ Usp.\  {\bf 39} (1996) 461]
  [arXiv:hep-ph/9603208].

\bibitem{Kajantie:1996mn}
  K.~Kajantie, M.~Laine, K.~Rummukainen and M.~E.~Shaposhnikov,
  Phys.\ Rev.\ Lett.\  {\bf 77} (1996) 2887
  [arXiv:hep-ph/9605288].

\bibitem{Shaposhnikov:1987pf}
  M.~E.~Shaposhnikov,
  Nucl.\ Phys.\  B {\bf 299} (1988) 797.

\bibitem{Gavela:1994ds}
  M.~B.~Gavela, M.~Lozano, J.~Orloff and O.~Pene,
  Nucl.\ Phys.\  B {\bf 430}, 345 (1994)
  [arXiv:hep-ph/9406288].

\bibitem{Gavela:1994dt}
  M.~B.~Gavela, P.~Hernandez, J.~Orloff, O.~Pene and C.~Quimbay,
  Nucl.\ Phys.\  B {\bf 430}, 382 (1994)
  [arXiv:hep-ph/9406289].

\bibitem{Fromme:2006cm}
  L.~Fromme, S.~J.~Huber and M.~Seniuch,
  JHEP {\bf 0611}, 038 (2006)
  [arXiv:hep-ph/0605242].

\bibitem{Carena:2002ss}
  M.~S.~Carena, M.~Quiros, M.~Seco and C.~E.~M.~Wagner,
  Nucl.\ Phys.\  B {\bf 650}, 24 (2003)
  [arXiv:hep-ph/0208043].

\bibitem{Prokopec:2003is}
  T.~Prokopec, K.~Kainulainen, M.~G.~Schmidt and S.~Weinstock,
  arXiv:hep-ph/0302192.

\bibitem{Carena:2004ha}
  M.~S.~Carena, A.~Megevand, M.~Quiros and C.~E.~M.~Wagner,
  Nucl.\ Phys.\  B {\bf 716}, 319 (2005)
  [arXiv:hep-ph/0410352].

\bibitem{Konstandin:2005cd}
  T.~Konstandin, T.~Prokopec, M.~G.~Schmidt and M.~Seco,
  Nucl.\ Phys.\  B {\bf 738}, 1 (2006)
  [arXiv:hep-ph/0505103].

\bibitem{Huber:2006wf}
  S.~J.~Huber, T.~Konstandin, T.~Prokopec and M.~G.~Schmidt,
  Nucl.\ Phys.\  B {\bf 757}, 172 (2006)
  [arXiv:hep-ph/0606298].

\bibitem{Chung:2008aya}
  D.~J.~H.~Chung, B.~Garbrecht, M.~J.~Ramsey-Musolf and S.~Tulin,
  Phys.\ Rev.\ Lett.\  {\bf 102}, 061301 (2009)
  [arXiv:0808.1144 [hep-ph]].

\bibitem{Jarlskog:1985ht}
  C.~Jarlskog,
  Phys.\ Rev.\ Lett.\  {\bf 55} (1985) 1039.


\bibitem{Salcedo:2000hp}
  L.~L.~Salcedo,
  Eur.\ Phys.\ J.\  C {\bf 20}, 147 (2001)
  [arXiv:hep-th/0012166].

\bibitem{Salcedo:2000hx}
  L.~L.~Salcedo,
  Eur.\ Phys.\ J.\  C {\bf 20}, 161 (2001)
  [arXiv:hep-th/0012174].

\bibitem{Konstandin:2003dx}
  T.~Konstandin, T.~Prokopec and M.~G.~Schmidt,
  Nucl.\ Phys.\  B {\bf 679}, 246 (2004)
  [arXiv:hep-ph/0309291].

\bibitem{Hernandez:2007ng}
  A.~Hernandez, T.~Konstandin and M.~G.~Schmidt,
  Nucl.\ Phys.\  B {\bf 793}, 425 (2008)
  [arXiv:0708.0759 [hep-th]].

\bibitem{Smit:2004kh}
  J.~Smit,
  JHEP {\bf 0409}, 067 (2004)
  [arXiv:hep-ph/0407161].

\bibitem{GarciaRecio:2009zp}
  C.~GarciaRecio and L.~L.~Salcedo,
  JHEP {\bf 0907}, (2009) 015
  [arXiv:0903.5494 [hep-ph]].


\bibitem{GarciaBellido:1999sv}
  J.~Garcia-Bellido, D.~Y.~Grigoriev, A.~Kusenko and M.~E.~Shaposhnikov,
  Phys.\ Rev.\  D {\bf 60} (1999) 123504
  [arXiv:hep-ph/9902449].

\bibitem{Krauss:1999ng}
  L.~M.~Krauss and M.~Trodden,
  Phys.\ Rev.\ Lett.\  {\bf 83} (1999) 1502
  [arXiv:hep-ph/9902420].

\bibitem{Copeland:2001qw}
  E.~J.~Copeland, D.~Lyth, A.~Rajantie and M.~Trodden,
  Phys.\ Rev.\  D {\bf 64}, 043506 (2001)
  [arXiv:hep-ph/0103231].

\bibitem{Turok:1990in}
  N.~Turok and J.~Zadrozny,
  Phys.\ Rev.\ Lett.\  {\bf 65} (1990) 2331.

\bibitem{Tranberg:2003gi}
  A.~Tranberg and J.~Smit,
  JHEP {\bf 0311} (2003) 016
  [arXiv:hep-ph/0310342].

\bibitem{vanTent:2004rc}
  B.~J.~W.~van Tent, J.~Smit and A.~Tranberg,
  JCAP {\bf 0407} (2004) 003
  [arXiv:hep-ph/0404128].


\bibitem{vanderMeulen:2005sp}
  M.~van der Meulen, D.~Sexty, J.~Smit and A.~Tranberg,
  JHEP {\bf 0602} (2006) 029
  [arXiv:hep-ph/0511080].

\bibitem{Tranberg:2006ip}
  A.~Tranberg and J.~Smit,
  JHEP {\bf 0608} (2006) 012
  [arXiv:hep-ph/0604263].

\bibitem{Tranberg:2006dg}
  A.~Tranberg, J.~Smit and M.~Hindmarsh,
  JHEP {\bf 0701} (2007) 034
  [arXiv:hep-ph/0610096].


\bibitem{Enqvist:2010fd}
  K.~Enqvist, P.~Stephens, O.~Taanila and A.~Tranberg,
  arXiv:1005.0752 [astro-ph.CO].

\bibitem{Konstandin:2011ds}
  T.~Konstandin and G.~Servant,
  arXiv:1104.4793 [hep-ph].



\bibitem{Felder:2000hj}
  G.~N.~Felder, J.~Garcia-Bellido, P.~B.~Greene, L.~Kofman, A.~D.~Linde and I.~Tkachev,
  Phys.\ Rev.\ Lett.\  {\bf 87} (2001) 011601
  [arXiv:hep-ph/0012142].


\bibitem{Rajantie:2000nj}
  A.~Rajantie, P.~M.~Saffin and E.~J.~Copeland,
  Phys.\ Rev.\  D {\bf 63}, 123512 (2001)
  [arXiv:hep-ph/0012097].

\bibitem{GarciaBellido:2002aj}
  J.~Garcia-Bellido, M.~Garcia Perez and A.~Gonzalez-Arroyo,
  Phys.\ Rev.\  D {\bf 67} (2003) 103501
  [arXiv:hep-ph/0208228].

\bibitem{Smit:2002yg}
  J.~Smit and A.~Tranberg,
  JHEP {\bf 0212}, 020 (2002)
  [arXiv:hep-ph/0211243].

\bibitem{GarciaBellido:2003wd}
  J.~Garcia-Bellido, M.~Garcia-Perez and A.~Gonzalez-Arroyo,
  Phys.\ Rev.\  D {\bf 69} (2004) 023504
  [arXiv:hep-ph/0304285].

\bibitem{Skullerud:2003ki}
  J.~I.~Skullerud, J.~Smit and A.~Tranberg,
  JHEP {\bf 0308}, 045 (2003)
  [arXiv:hep-ph/0307094].

\bibitem{DiazGil:2007dy}
  A.~Diaz-Gil, J.~Garcia-Bellido, M.~Garcia Perez and A.~Gonzalez-Arroyo,
  Phys.\ Rev.\ Lett.\  {\bf 100} (2008) 241301
  [arXiv:0712.4263 [hep-ph]].

\bibitem{Ambjorn:1990pu}
  J.~Ambjorn, T.~Askgaard, H.~Porter and M.~E.~Shaposhnikov,
  Nucl.\ Phys.\  B {\bf 353}, 346 (1991)

\bibitem{Bodeker:1999gx}
  D.~Bodeker, G.~D.~Moore and K.~Rummukainen,
  Phys.\ Rev.\  D {\bf 61} (2000) 056003
  [arXiv:hep-ph/9907545].

\bibitem{WMAP}
  N.~Jarosik {\it et al.},
  arXiv:1001.4744 [astro-ph.CO].

\end{thebibliography}
\end{document}